\documentclass[showpacs,amsmath,amssymb]{revtex4}

\usepackage{graphicx}
\usepackage{dcolumn}
\usepackage{bm}
\usepackage{epsfig}

\begin{document}

\def\0#1#2{\frac{#1}{#2}}
\def\bct{\begin{center}} \def\ect{\end{center}}
\def\beq{\begin{equation}} \def\eeq{\end{equation}}
\def\bea{\begin{eqnarray}} \def\eea{\end{eqnarray}}
\def\nnu{\nonumber}
\def\n{\noindent} \def\pl{\partial}
\def\g{\gamma}  \def\O{\Omega} \def\e{\varepsilon} \def\o{\omega}
\def\s{\sigma}  \def\b{\beta} \def\p{\psi} \def\r{\rho}
\def\G{\Gamma} \def\S{\Sigma} \def\l{\lambda}

\title{The vacuum tunnelling and the crossover of deconfinement in Friedberg-Lee model}
\author{Song~Shu}
\affiliation{Faculty of Physics and Electronic Technology, Hubei
University, Wuhan 430062, China}
\author{Jia-Rong~Li}
\affiliation{Institute of Particle Physics, Hua-Zhong Normal
University, Wuhan 430079, China} \affiliation{Key Laboratory of
Quark {\rm\&} Lepton Physics (Hua-Zhong Normal University),
Ministry of Education, China}
\begin{abstract}
We have discussed the vacuum tunnelling in Friedberg-Lee model.
The tunnelling coefficient is derived in the field configuration
space by calculating the transition amplitude using the path
integral under the stationary phase approximation and the dilute
instanton gas approximation. By studying the tunnelling effect
between the two degenerating vacuums at the critical temperature
and chemical potential, we find that the system could be
deconfined by tunnelling, which could possibly change the first
order deconfinement phase transition to crossover.
\end{abstract} \pacs{25.75.Nq, 12.39.Ki, 11.10.Wx, 03.65.Xp} \maketitle

\section{Introduction}
QCD Lagrangian has two remarkable symmetries. One is
$SU(N_f)_L\times SU(N_f)_R$ chiral symmetry at zero quark mass
limit, which is global symmetry in flavor space. The other is
color $SU(N_c)$ symmetry, which is the local symmetry in color
space. At infinite quark mass limit, it is $Z(N_c)$ center
symmetry. The chiral phase transition is related to the chiral
symmetry, which is characterized by the chiral condensate, while
the deconfinement phase transition is related to the $Z(N_c)$
center symmetry, which is characterized by the Polyakov
loop~\cite{ref1,ref2}. Theoretically these two phase transitions
have been mixed up while discussing the phase structure of QCD
phase diagram. The strict and reliable theoretical calculation
with full QCD dynamics at present is lattice
calculation~\cite{ref3,ref4,ref5}. However as the two symmetries
are in the color space and flavor space separately, in model
studying these two symmetries can be studied separately by two
different types of models. One is the chiral model, such as
Nambu-Jona-Lasinio (NJL) model, linear sigma model,
etc~\cite{ref6,ref7,ref8,ref9,ref10,ref11}. The other is the bag
model, such as MIT model~\cite{ref12}, SLAC model~\cite{ref13} and
Friedberg-Lee(FL) model~\cite{ref14}. Until recent years when
Polyakov loop was combined into the chiral model, such as PNJL
model and Polyakov loop linear sigma
model~\cite{ref15,ref16,ref17,ref18,ref19}, one starts to
investigate both chiral phase transition and deconfinement phase
transition in the same model. However the results are
qualitatively consistent with that of the Lattice. So far by
summing up the results of both lattice calculation and model
studying one can see that in the $T-\mu$ phase diagram at high
temperatures and low densities there exists a crossover from
nuclear matter to quark-gluon plasm (QGP), while at low
temperatures and high densities it is a first order phase
transition. With temperature increasing and density decreasing
from first order phase transition to crossover there exists a
critical end point (CEP)~\cite{ref3,ref20,ref21}.

Within the context of the bag model, the first order transition is
a natural consequence. As the quarks are confined in a hadron,
there is huge binding energy. From hadronic phase to quark phase,
there should be enormous latent heat, which will result in a first
order phase transition. What is difficult to understand is the
crossover, because that means the transformation between the
hadronic phase and the quark phase will be smooth. Thus there will
be no latent heat at all. From the theoretical point of view, what
is the physical mechanism of the crossover? In this paper we focus
our study on deconfinement of the system, and we want to make a
tentative study of the possible mechanism of the crossover of the
deconfinement.

In the earlier studies the intuitive physical picture of
deconfinement is that by compressing the system of nucleons
through colliding nuclei, the quarks can move freely among
different nucleons when the nucleons described by the wave
functions become overlap. This kind of picture were further
developed by percolation model~\cite{ref22}. With increasing
temperatures and densities, the quarks can percolate among
different hadrons by bounds to form clusters and eventually become
QGP. However in this picture it is lack of dynamics to describe
the percolation. In Ref.~\cite{ref23}, it is assumed that as a
single hadron the quarks are confined in a bag produced by an
infinite deep harmonic potential well. When the two hadrons get
close, the infinite potential barrier between the neighboring
wells becomes finite. Thus the quarks in adjacent hadrons are
allowed to penetrate through the finite potential barrier. However
these studies are limited to the quantum mechanical level. In this
paper we want to make the discussions based on quantum field
theory. FL model is a bag model based on quantum field theory,
which contains the essential features of color confinement of
QCD~\cite{ref14}. It is a covariant Lagrangian formulation by
which the dynamics of the confinement could be handled
conveniently. At different approximations, the FL model could
reproduce the results of both SLAC model and MIT model, and it has
been very successful in describing phenomenologically the static
properties of hadrons and their behaviors at low
energies~\cite{ref24,ref25,ref26}. Especially, it could be
extended to finite temperature field theory and deconfinement
phase transition could be studied at finite temperatures and
densities. FL model is most suitable for us to carry out our
discussions in this paper, so we choose FL model to fulfill our
goal.

The most remarkable feature of FL model is the dielectric property
of the vacuum~\cite{ref27}. T.D. Lee indicated that vacuum
resembled a medium which properties could be
changed~\cite{ref27a}, and he further proposed ``vacuum
engineering'' which could be realized by high energy collisions of
heavy nuclei. Vacuum has a complex structure. In FL model a
phenomenological scalar field $\s$, which can be identified with
the gluon condensate arising from the nonlinear interactions of
color fields, is introduced to describe the complex structure of
the vacuum. There are two vacuums in this model. One is the
physical vacuum or the non perturbative vacuum in which the $\s$
field attains a nonzero vacuum value $\s_v$. And the physical
vacuum may well be a perfect color anti-dielectric substance. The
other is perturbative vacuum $\s=0$ in which the quarks are nearly
free and perturbative QCD is assumed to be valid. In the absence
of quarks, the normal state of the $\s$ field is at the vacuum
value $\s_v$. In the presence of quarks, the $\s$ field finds a
minimum in the vicinity of zero. That means the quarks dig a hole
in the physical vacuum, which results in a soliton bag
corresponding to a hadron. This is the origin of confinement in
the model. The model has been also extended to finite temperatures
and densities to study deconfinement phase
transition~\cite{ref28,ref29,ref30,ref31,ref32}. In the previous
studies, the deconfinement phase transition had been often
discussed by analyzing the changing of the two vacuums. The energy
difference between the two vacuums defines the bag constant $B$.
When the temperature or density increases, the energy difference
between the two vacuums decreases. Until the two vacuums
degenerate, the bag constant becomes to zero, and the
deconfinement phase transition takes place, which is a first order
phase transition.

However the tunnelling effect between the two degenerate vacuums
are not considered in these studies. It has been indicated by
T.~D.~Lee that it is an interesting problem to consider the
tunnelling effect between the two degenerate vacuums in field
configuration space in a finite size microscopic
system~\cite{ref27}. Here we should make a special emphasis on the
meaning of the tunnelling in our study. The tunnelling here is the
tunnelling between the two degenerate vacuums rather than the
tunnelling of quarks among different bags. That means at the time
of deconfinement where the two vacuum degenerate, the perturbative
vacuum inside a hadron and the physical vacuum outside a hadron
could tunnel each other. Thus the tunnelling effect here is
completely independent of typical inter-particle distance among
baryons. We think that the tunnelling effect between the two
vacuums may be important, and it could be possible physical
mechanism for the crossover of the deconfinement.

The organization of this paper is as follows: in section 2 the FL
model is introduced. The field equation of the sigma field is
treated in homogeneous case. An effective lagrangian is obtained
from which a transition amplitude is given in a form of the path
integral. In section 3, the transition amplitude is evaluated in
Euclidean space in stationary phase approximation (SPA) and dilute
instanton gas approximation. The tunnelling amplitude is further
derived in field configuration space. In section 4, the tunnelling
probability is evaluated for the different effective potentials at
different critical temperatures and chemical potentials. The
tunnelling effect is studied in the deconfinement phase
transition. The last section is the summary.

\section{The effective Lagrangian and the transition amplitude of FL model}
We start from the Lagrangian of the FL model, \bea {\cal
L}=\bar\psi(i\gamma_\mu\pl^\mu-g\s)\psi+\012(\pl_\mu\s)(\pl^\mu\s)-U(\s),
\eea where\bea U(\s)=\01{2!}a\s^2+\01{3!}b\s^3+\01{4!}c\s^4+B.
\eea $\p$ represents the quark field, and $\s$ denotes the
phenomenological scalar field. $a, b, c, g$ and $B$ are the
constants which are generally fitted in with producing the
properties of hadrons appropriately at zero temperature. We shift
the $\s$ field as $\s\rightarrow\s+\s'$ where the new $\s$ and
$\s'$ are the vacuum expectation value and the fluctuation of the
$\s$ field respectively. Then the lagrangian becomes \bea {\cal
L}^\prime=\bar\psi(i\gamma_\mu\pl^\mu-m_q)\psi+\012(\pl_\mu\s')(\pl^\mu\s')-\012m_\s^2\s'^2-U(\s),
\eea where \bea U(\s)=\01{2!}a\s^2+\01{3!}b\s^3+\01{4!}c\s^4+B.
\eea $m_q=g\s$ and $m_\s^2=a+b\s+\012c\s^2$ are the effective
masses of the quark and $\s$ field respectively. The interactions
associated with the fluctuation $\s'$, such as $\s'^3$, $\s'^4$
and $\bar\psi\s'\psi$, are neglected in the usual tree level
approximation. At the tree level, one can obtain the field
equations, \beq (i\gamma_\mu\pl^\mu-g\s)\psi=0, \eeq \beq
\pl_\mu\pl^\mu\s'+m_\s^2\s'=0, \eeq \beq
\pl_\mu\pl^\mu\s=-\left(\0{\pl
U}{\pl\s}+\012(b+c\s)\langle{\s^\prime}^2\rangle+g\langle\bar\psi\psi\rangle\right)\equiv-\0{\pl
V_{eff}(\s)}{\pl\s}, \label{mean} \eeq where
$\langle{\s'}^2\rangle$ and $\langle\bar\psi\psi\rangle$ denote
the contributions of the thermal excitations of $\s$ and quark
fields respectively. The second equality of equation~(\ref{mean})
means that we define a thermal effective potential. These thermal
functions will be determined in the later section. Here our
discussion will focus on the equation~(\ref{mean}). $\s$ is the
classic field which is time and space dependent. There are two
physical treatments of $\s$. If one treats the $\s$ time
independent and space dependent which means the static case, the
equation~(\ref{mean}) will be the usual soliton equation which is
solved for soliton solutions. This case is mostly studied for the
confinement mechanics and hadron properties in the previous
literature. If one treats the $\s$ time dependent and space
independent which means the homogeneous case. In this case the
solution of the equation will be the instanton which is related to
the tunnelling effect. Here we are interested in tunnelling
effect, so we take the homogeneous case and the equation becomes
\bea \0{d^2\s}{dt^2}=-\0{\pl V_{eff}(\s)}{\pl\s}. \eea We can
regard the $\s$ field as a ``space coordinate" of a particle
moving in one dimension in a potential $V_{eff}(\s)$. By this
analog we can further obtain the effective Lagrangian density
which describes the motion of the particle, \beq {\cal
L}_{eff}=\012\left(\0{d\s}{dt}\right)^2-V_{eff}(\s). \eeq For the
homogeneous case the effective Lagrangian is \beq L_{eff}=\int
d^3x{\cal
L}_{eff}=\O\left[\012\left(\0{d\s}{dt}\right)^2-V_{eff}(\s)\right],
\eeq in which $\O$ is the space volume. The effective Lagrangian
could be quantized by the canonical quantization. The
corresponding quantized Hamiltonian in quantum theory is \beq
H=\0{\hat{p}^2}{2\O}+\O V_{eff}(\s), \eeq where $\hat
p=-i\hbar\pl/\pl\s$ is the canonical momentum operator.
Consequently the quantum transition amplitude for a particle to
start at ``position" $\s=\s_a$ at time $t=0$, and end up at
$\s=\s_b$ at $t=T$ is given by \beq
K(\s_b,T;\s_a,0)=<\s_b|exp(-iHT/\hbar)|\s_a>. \label{prog}\eeq By
the path integral the transition amplitude can be also written as
\beq K(\s_b,T;\s_a,0)=\int D[\s(t)]exp \{i/\hbar S[\s(t)]\}, \eeq
where \beq S[\s(t)]=\int_0^TdtL_{eff}=\int_0^Tdt\O\left
[\012\left(\0{d\s}{dt}\right)^2-V_{eff}(\s)\right], \eeq is the
action and $\int D[\s(t)]$ is an integral over all paths.

\section{Instanton and tunnelling effect}
\begin{figure}[tbh]
\begin{center}
\includegraphics[width=210pt,height=150pt]{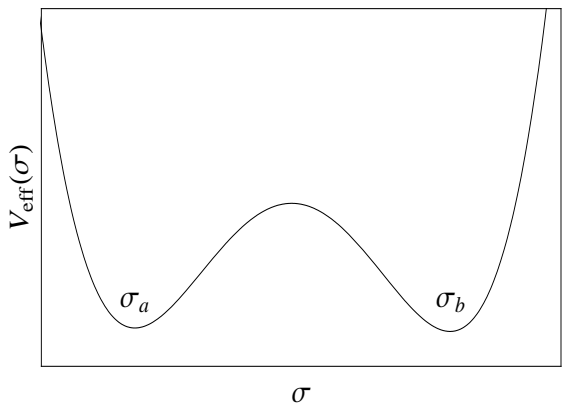}
\end{center}
\caption{The schematically sketched thermal effective potential as
a function of $\s$. }\label{f1}
\end{figure}

In order to evaluate the transition amplitude, it is more
convenient to do the calculation in a Euclidean space. The
corresponding Euclidean transition amplitude denotes as \beq
K_E(\s_b,\tau/2;\s_a,-\tau/2)=\int D_E[\s(\tau')]exp[-1/\hbar
S_E(\s(\tau')], \label{trans}\eeq where \beq
S_E[\s(\tau')]=\int_{-\tau/2}^{\tau/2}d\tau'\O\left[\012\left(\0{d\s}{d\tau'}\right)^2+V_{eff}(\s)\right]
\eeq is the Euclidean action, $\tau=iT$ and $\tau'=it$. One can
also obtain the Euclidean equation of motion, \beq
\0{d^2\s}{d\tau'^2}=\0{\pl V_{eff}(\s)}{\pl\s}. \label{inst} \eeq
In the following discussion we take $\tau\rightarrow\infty$ and we
suppose $V_{eff}(\s)$ has the configuration schematically sketched
in Fig.\ref{f1}.
\begin{figure}[tbh]
\begin{center}
\includegraphics[width=210pt,height=150pt]{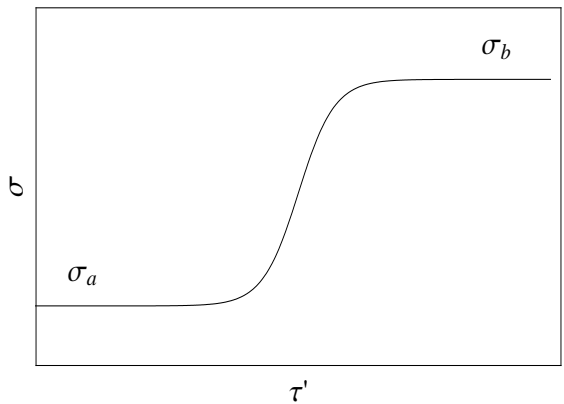}
\end{center}
\caption{A single instanton solution in the FL model.}\label{f2}
\end{figure}
The potential has two minima which are degenerate. The
equation~(\ref{inst}) could be solved by the boundary condition
\beq \tau'=-\infty, \ \s=\s_a; \ \ \ \ \ \ \tau'=\infty, \
\s=\s_b. \label{bond}\eeq  The solution is an instanton which is
schematically sketched in Fig.\ref{f2}. This instanton solution
$\s_{cl}(\tau')$ is a classical path which is an extremum of the
action $S_E[\s(\tau')]$. One could expand the action
$S_E[\s(\tau')]$ about the classical path $\s=\s_{cl}(\tau')$,
\beq
S_E[\s(\tau')]=S_E[\s_{cl}(\tau')]+\0{\O}{2}\int_{-\infty}^{\infty}d\tau'y(\tau')O_E(\tau')y(\tau')+O(y^3),
\label{action} \eeq where \beq y(\tau')=\s(\tau')-\s_{cl}(\tau'),
\eeq and \beq
O_E(\tau')=-\0{\pl^2}{\pl\tau'^2}+\left(\0{\pl^2V_{eff}(\s)}{\pl\s^2}\right)_{\s_{cl}(\tau')}.
\eeq In the spirit of the stationary phase approximation (SPA),
the major contribution to the transition amplitude (\ref{trans})
will come from the vicinity of the classical path
$\s(\tau')=\s_{cl}(\tau')$. In that case, to leading approximation
one can neglect the cubic and higher terms in $y(\tau')$ in the
Taylor series (\ref{action}). The SPA as applied to the path
integral in the transition amplitude gives \beq
K_E(\s_b,\tau/2;\s_a,-\tau/2)=e^{-S_0/\hbar}B_E(\tau)\left[DetO_E(\tau')\right]^{-\012},
\label{kdet} \eeq where $S_0=S_E[\s_{cl}(\tau')]$ and
$B_{E}(\tau)$ is the measure factor in functional integral. $Det$
is the determinant of the operator $O_E(\tau')$ which is most
conveniently obtained from its eigenvalues, but the operator will
have a zero eigenvalue because the action is translation invariant
in the $\tau'$ variable. Formally this will cause a divergence in
(\ref{kdet}). However this zero mode could be separated from the
determinant. For a detail discussion of this treatment one could
refer to~\cite{ref33}. After separating the zero mode the result
is \beq K_E(\s_b,\tau/2;\s_a,-\tau/2)=e^{-S_0/\hbar}J\tau
B_E(\tau)\left[Det'O_E(\tau')\right]^{-\012}, \label{kdet1}\eeq
where $J=\sqrt{S_0}$ and $Det'$ denotes the determinant without
the zero mode. In order to calculate this determinant, one could
be referred to the case of harmonic oscillator. If the particle
oscillates in the vicinity of the ``position" $\s_a$, it could be
regarded as a harmonic oscillator. So is it in the vicinity of the
``position" $\s_b$. We suppose the angular frequency of the
harmonic oscillator is $\o$. The determinant can be written as
\beq
\left[Det'O_E(\tau')\right]^{-\012}=\left[Det(-\0{d^2}{d\tau'^2}+\o^2)\right]^{-\012}k,
\eeq where $k$ is a constant independent of $\tau$ as
$\tau\rightarrow\infty$. For the harmonic oscillator, the
determinant could be evaluated at $\tau\rightarrow\infty$ and the
result is \beq
B_E(\tau)\left[Det(-\0{d^2}{d\tau'^2}+\o^2)\right]^{-\012}=\left(\0{\o}{\pi\hbar}\right)^\012e^{-\0{\o\tau}2}.
\eeq Hence the amplitude (\ref{kdet1}) becomes \beq
K_E(\s_b,\tau/2;\s_a,-\tau/2)=e^{-S_0/\hbar}Jk\tau\left(\0{\o}{\pi\hbar}\right)^\012e^{-\0{\o\tau}2}.
\eeq

\begin{figure}[tbh]
\begin{center}
\includegraphics[width=210pt,height=150pt]{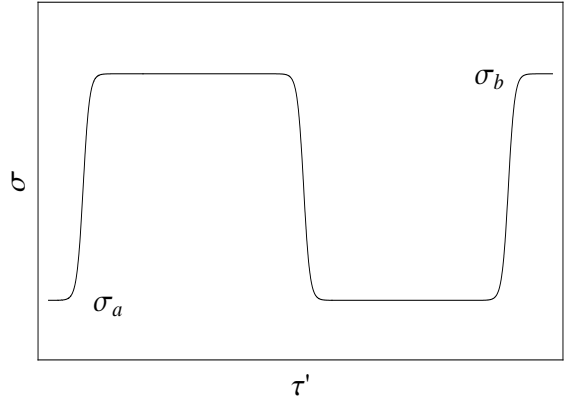}
\end{center}
\caption{Multi-instanton solution in the FL model.}\label{f3}
\end{figure}

From the above derivation we obtain the result of the transition
amplitude by considering single instanton. From the notion of
tunnelling, this means the particle traverses the potential
barrier once. However the particle could also traverse the
potential barrier several times, that means there are also
multi-instanton solution which configuration is shown in
Fig.\ref{f3} that will have the correct boundary condition for our
problem, namely $\s(-\infty)=\s_a$, $\s(\infty)=\s_b$.
The instantons should be widely separated and is so called the
instanton gas. For our interest of tunnelling effect, the
contributions from all these multi-instanton solutions to the
transition amplitude should be summed up. The summation could be
evaluated under the dilute instanton-gas
approximation~\cite{ref33} and the result is \beq
K_E(\s_b,\tau/2;\s_a,-\tau/2)=\left(\0{\o}{\pi\hbar}\right)^\012e^{-\0{\o\tau}2}sinh\left(Jk\tau
e^{-S_0/\hbar}\right). \eeq By equation~(\ref{prog}) the
transition amplitude can be also written as \beq
<\s_b|exp(-H\tau/\hbar)|\s_a>=\left(\0{\o}{\pi\hbar}\right)^\012e^{-\0{\o\tau}2}sinh\left(Jk\tau
e^{-S_0/\hbar}\right). \label{s1} \eeq By the derivation in a
similar way to the above derivation one can also obtain other
three kinds of transition amplitudes in the following, \bea
<\s_a|exp(-H\tau/\hbar)|\s_b>=\left(\0{\o}{\pi\hbar}\right)^\012e^{-\0{\o\tau}2}sinh\left(Jk\tau
e^{-S_0/\hbar}\right), \label{s2} \\
<\s_a|exp(-H\tau/\hbar)|\s_a>=\left(\0{\o}{\pi\hbar}\right)^\012e^{-\0{\o\tau}2}cosh\left(Jk\tau
e^{-S_0/\hbar}\right), \label{s3} \\
<\s_b|exp(-H\tau/\hbar)|\s_b>=\left(\0{\o}{\pi\hbar}\right)^\012e^{-\0{\o\tau}2}cosh\left(Jk\tau
e^{-S_0/\hbar}\right). \label{s4} \eea If we define the states
\beq |\s_\pm>=\01{\sqrt{2}}(|\s_a>\pm|\s_b>), \eeq then from
equations~(\ref{s1}---\ref{s4}) we have \beq
<\s_{\pm}|exp(-H\tau/\hbar)|\s_{\pm}>=\left(\0{\o}{\pi\hbar}\right)^\012e^{-\0{\o\tau}2\pm
Jk\tau e^{-S_0/\hbar}}. \eeq From the above expression one could
read out the energy \beq E_\pm=\012\hbar\o\mp\hbar
Jke^{-S_0/\hbar}, \label{e} \eeq from which one could see that the
states $|\s_{\pm}>$ are the eigen-states of $H$ with the
eigenvalues $E_\pm$. The first term in equation (\ref{e}) is the
energy without tunnelling while the second term is the energy
correction coming from the contribution of the tunnelling through
the potential barrier. The factor $e^{-S_0/\hbar}$ reveals the
exponential dependence on the ``barrier strength". Notice that
\beq
S_0=S_E[\s_{cl}(\tau')]=\int_{-\infty}^{\infty}d\tau'\O\left[\012\left(\0{d\s}{d\tau'}\right)^2+V_{eff}(\s)\right]_{\s_{cl}(\tau')},
\label{s0} \eeq in which $\s_{cl}(\tau')$ is the classical path
and will obey \beq \012\left(\0{d\s}{d\tau'}\right)^2=V_{eff}(\s).
\label{inst1}\eeq The above equation can be derived from the
equation~(\ref{inst}) by integration. Substituting
equation~(\ref{inst1}) into equation~(\ref{s0}) and changing the
integration variable we obtain \beq
S_0=\int_{\s_a}^{\s_b}\O\sqrt{2V_{eff}(\s)}d\s. \eeq This quantity
represents the strength of the potential barrier between $\s_a$
and $\s_b$. It is clear that the tunnelling amplitude is \beq
M=exp\left({-\0{S_0}{\hbar}}\right)=exp\left[{-\0{\O}{\hbar}\int_{\s_a}^{\s_b}\sqrt{2V_{eff}(\s)}d\s}\right].
\eeq The result is in accord with conventional WKB treatments of
quantum mechanics in the position configuration space, but one
should notice that the result here is in the field configuration
space.

\section{Crossover of deconfinement by tunnelling}
By the tunnelling amplitude one can further obtain the tunnelling
probability \beq
\eta=|M|^2=exp\left[-\0{2\O}{\hbar}\int_{\s_a}^{\s_b}\sqrt{2V_{eff}(\s)}d\s\right].
\label{tunl} \eeq It is clear that the tunnelling probability is
mainly dependent on the potential $V_{eff}(\s)$ and the volume of
the system which we will discuss later. Recall
equation~(\ref{mean}), the thermal excitations of $\s$ and quark
fields could be evaluated by standard method of finite temperature
field theory and the results are \bea \langle{\s'}^2\rangle &=&
\int\0{d^3\bf
p}{(2\pi)^3}\01{E_\s}\01{e^{\b E_\s}-1}, \label{sgm} \\
\langle\bar\psi\psi\rangle &=& -\g\int \0{d^3\bf
p}{(2\pi)^3}\0{m_q}{E_q}\left(\01{e^{\b(E_q-\mu)}+1}+\01{e^{\b(E_q+\mu)}+1}\right),
\label{psi} \eea in which $\b$ is the inverse temperature, $\mu$
is the chemical potential, and $\g$ is a degenerate factor,
$\g=2(spin)\times 2(flavor)\times 3(color)$. $E_\s=\sqrt{{\bf
p}^2+m_\s^2}$ and $E_q=\sqrt{{\bf p}^2+m_q^2}$. For the values of
the model parameters $a, b, c$ and $g$, there are different
choices. We have taken one set of values as $a=17.7fm^{-2},
b=-1457.4fm^{-1}, c=20000, g=12.16$, which have been often used in
the literature~\cite{ref30}. The different choice of the
parameters will not qualitatively change the results in our
following discussion. The effective mass of $\s$ field is fixed at
$m_\s=550MeV$. One should notice that the second equality of
equation~(\ref{mean}) means that we define a thermal effective
potential. By equation (\ref{sgm}) and (\ref{psi}) the thermal
effective potential is \bea V_{eff}(\s)=U(\s)+\01{\b}\int
\0{d^3\bf p}{(2\pi)^3}\ln (1-e^{-\b E_{\s}})-\0{\g}{\b}\int
\0{d^3\bf p}{(2\pi)^3}\left[\ln (1+e^{-\b(E_q-\mu)}) + \ln
(1+e^{-\b(E_q+\mu)})\right]. \label{eff} \eea

\begin{figure}[tbh]
\begin{center}
\includegraphics[width=210pt,height=150pt]{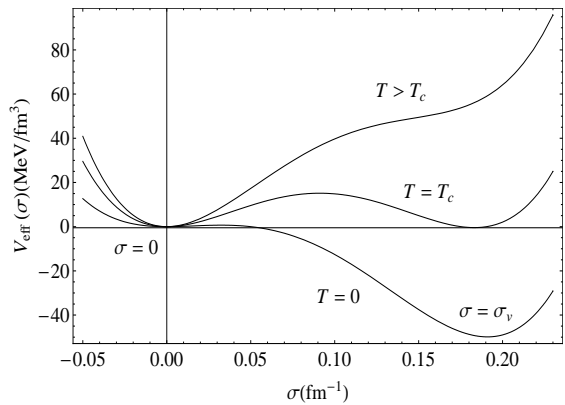}
\end{center}
\caption{The thermal effective potentials at the different
temperatures.}\label{f4}
\end{figure}

The effective potential could be numerical evaluated at the given
temperature and chemical potential. First let us make a brief
review of usually deconfinement phase transition in FL model at
finite temperature. At zero temperature, the $V_{eff}(\s)$ is just
the $U(\s)$ which has two minima corresponding to the two vacuums:
one is the perturbative vacuum at $\s=0$; the other is the
nonperturbative vacuum at $\s=\s_v\neq 0$, which is the absolute
minimum corresponding to the true vacuum. See Fig.\ref{f4}. When
the system is heated up, the nonperturbative vacuum is lifted up,
and the energy difference between the two vacuums decreases. Until
the critical temperature $T_c$ that the two vacuums degenerate,
the deconfinement phase transition takes place. Keep heating the
system, which means providing the latent heat, the nonperturbative
vacuum $\s=\s_v$ will become metastable. At some higher
temperature $T>T_c$, the nonperturbative vacuum $\s=\s_v$
disappears and the perturbative vacuum $\s=0$ becomes the unique
minimum corresponding to the true vacuum. The system undergoes a
first order deconfinement phase transition. This scenario is the
deconfinement phase transition without tunnelling.

Now we are going to discuss the tunnelling effect. When the system
is heated to the critical temperature where the two vacuums
degenerate, the tunnelling may take place. If the tunnelling
intervenes, there is no need to keep on heating the system. By
tunnelling the nonperturbative vacuum $\s=\s_v$ will be smoothly
transferred to the perturbative vacuum without any energy changing
and there will be no latent heat as a result. The first order
phase transition will be replaced by a smooth crossover.

\begin{figure}[tbh]
\begin{center}
\includegraphics[width=210pt,height=150pt]{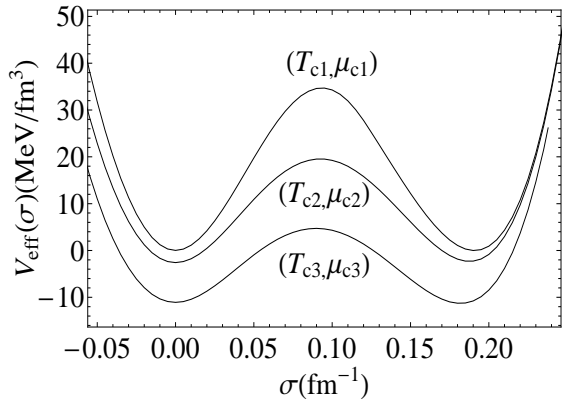}
\end{center}
\caption{The thermal effective potentials at the different
critical temperatures and chemical potentials. From top to bottom
the critical temperatures and chemical potentials are:
$T_{c1}=0MeV$, $\mu_{c1}=297MeV$; $T_{c2}=76MeV$,
$\mu_{c2}=210MeV$; $T_{c3}=121MeV$ $\mu_{c3}=0MeV$.}\label{f5}
\end{figure}

For further details of the tunnelling effect in deconfinement, we
need to study the tunnelling probability. The tunnelling
probability depends on the effective potential. From equation
(\ref{eff}), the effective potential could be evaluated at finite
temperatures and chemical potentials. When the two vacuums
degenerate we could obtain the critical temperature and critical
chemical potential. In Fig.\ref{f5} we show the thermal effective
potentials at the different critical temperatures and chemical
potentials. As we know it is easier for tunnelling to occur in
flat and narrow potential barriers than in tall and wide potential
barriers. From Fig.\ref{f5} we can see the potential barriers
become flatter and narrower from the top one to the bottom one,
which means, by a qualitative analysis, the tunnelling probability
becomes larger and it is easier for tunnelling to occur from the
top one to the bottom one. In order to make it more accurate we
will evaluate the tunnelling probability.

From equation~(\ref{tunl}) we could see that there is a volume
factor in the exponent. The tunnelling from one vacuum to another
vacuum is strongly suppressed when the volume of the system goes
to infinity. As in relativistic heavy ion collisions, the basic
ingredients are nucleons inside the nuclei. Here we replace the
size of the uniform vacuum domain with the size of a free nucleon.
In our calculation, we take $\hbar=1$ and $\O=4\pi R^3/3$. $R$ is
the radius of a free nucleon and the value is $R=1fm$. From
equation~(\ref{eff}) we could evaluate the thermal effective
potential at the critical temperature and chemical potential. By
equation~(\ref{tunl}) the tunnelling probability could be further
determined numerically. The tunnelling probabilities for different
critical temperatures and chemical potentials are shown in
Table~\ref{t1}.
\begin{table}
\caption{\label{t1}The tunnelling probabilities at the different
critical temperatures and chemical potentials.}
\begin{ruledtabular}
\begin{tabular}{ccc}
$\mu_c(MeV)$ & $T_c(MeV)$ & $\eta$ \\
\hline
0 & 121 & $66.3\%$\\
100 & 110 & $65.7\%$\\
150 & 97 & $64.8\%$\\
200 & 80 & $61.0\%$\\
210 & 76 & $59.9\%$\\
250 & 54 & $58.5\%$\\
297 & 0  & $55.2\%$\\
\end{tabular}
\end{ruledtabular}
\end{table}
From the table we could find that the tunnelling probability
increases with the critical temperature increasing (or with the
critical chemical potential decreasing). That is to say along the
phase boundary in the the $T-\mu$ phase diagram it is more
favorable for the tunnelling to occur at high critical
temperatures and low critical chemical potentials. Here the
nucleon size is fixed and we do not consider the variation of the
nucleon size with the temperature or density. The size variation
of the system may have important corrections to our results. To
determine the nucleon size at different temperatures and densities
in FL model is out of the scope of this work. However in present
work we could just give an estimation about the size dependent of
the tunnelling probability. By simply increasing the radius of the
system to $R=1.2fm$ we find that the tunnelling probability at the
point $\mu_c=0MeV$ and $T_c=121MeV$ will decrease to
$\eta=49.2\%$. Compared to the case of $R=1fm$ with the tunnelling
rate $\eta=66.3\%$ at the same point, the tunnelling rate for
$R=1.2fm$ have been decreased by $17.1\%$, which is indeed
remarkable. In a more serious consideration, there should be more
systematic and strict treatments and calculations about the
nucleon size effect, which deserve thorough investigations in the
future studies.

\begin{figure}[tbh]
\begin{center}
\includegraphics[width=210pt,height=150pt]{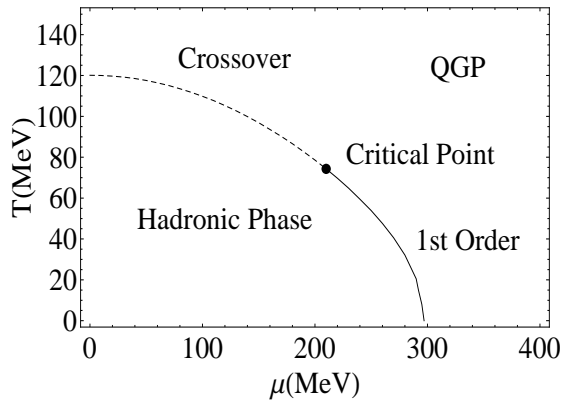}
\end{center}
\caption{$T-\mu$ phase diagram of deconfinement in the FL
model.}\label{f6}
\end{figure}

The $T-\mu$ phase diagram of deconfinement phase transition is
plotted in Fig.\ref{f6}. We can suppose that with the tunnelling
probability increasing along the phase boundary, at some critical
point, say $\eta\approx 60\%$, the tunnelling will play dominant
role and the system will be deconfined by tunnelling. After that
critical point along the phase boundary the first order phase
transition will be replaced by the crossover, as shown in
Fig.\ref{f6}. The critical point here is not determined in a
strict sense. It is obtained by the qualitative physical analysis.
Its location is not unique along the phase boundary. However in
our work we only wish to emphasize that the tunnelling could take
remarkable effect when the tunnelling probability is large enough,
which will change the nature of the phase transition and lead to
crossover. This physical picture is qualitatively consistent with
the present phase diagram of the QCD phase transition. The
tunnelling could be a possible physical mechanism of the crossover
of the deconfinement.

\section{summary}
In this paper we have discussed the crossover of deconfinement by
tunnelling in FL bag model. By the method of path integral we have
derived the tunnelling amplitude in the field configuration space
and obtain the tunnelling probability which is accord with the
conventional WKB treatment of quantum mechanics. The tunnelling
probabilities for the different thermal effective potentials in
field configuration space at the different critical temperatures
and chemical potentials are evaluated. In the context of the bag
model, we find that the tunnelling is more likely to take place at
the high critical temperatures and low critical chemical
potentials. We indicate that at some critical temperatures and
chemical potentials the system could be deconfined by tunnelling
process which will lead to a crossover instead of a first order
phase transition of deconfinement. However it should be reminded
that in this paper we only give one possible physical mechanism
which could lead to the crossover of deconfinement within the
context of bag model, and the corresponding numerical results on
the phase diagram are model dependent.

\begin{acknowledgments}
This work was supported in part by the National Natural Science
Foundation of China with No. 10905018 and No. 11275082.
\end{acknowledgments}

\end{document}